\documentclass{iau} 
\usepackage{graphicx}
\usepackage{natbib}
\usepackage{hyperref}
\usepackage{wrapfig}
\usepackage{sidecap}

\newcommand{\apj}{{\it Astrophys. J.}}

\newcommand{\aj}{{\it Astron. J.}}
\newcommand{\mnras}{{\it Mon. Not. R. Astron. Soc.}}

\newcommand{\aap}{{\it Astron. Astrophys.}}

\newcommand{\pasa}{{\it Publications of the Astron. Soc. of Australia}}

\title[JD 11.~~Machine Learning in Astronomy] 
{Pushing the Technical Frontier: \\ From Overwhelmingly Large Data Sets \\ to Machine Learning}

\author[V. Acquaviva]   
{Viviana Acquaviva$^1$}

\affiliation{$^1$Physics Department, New York City College of Technology, \\ 300 Jay Street, Brooklyn NY 11201 \\email: {\tt vacquaviva@citytech.cuny.edu}}

\pubyear{2019}
\volume{341}  
\jname{Challenges in Panchromatic Galaxy Modelling with Next Generation Facilities}
\editors{M. Boquien, E. Lusso, C. Gruppioni, and P. Tissera}
\begin{document}

\maketitle

\begin{abstract}
This paper summarizes my thoughts, given in an invited review at the IAU symposium 341 ``Challenges in Panchromatic Galaxy Modelling with Next Generation Facilities", about how machine learning methods can help us solve some of the big data problems associated with current and upcoming large galaxy surveys. 

\end{abstract}

\firstsection 
\section{It's everywhere!} 
\begin{figure}[b]
\begin{center}
\includegraphics[width=0.4 \textwidth]{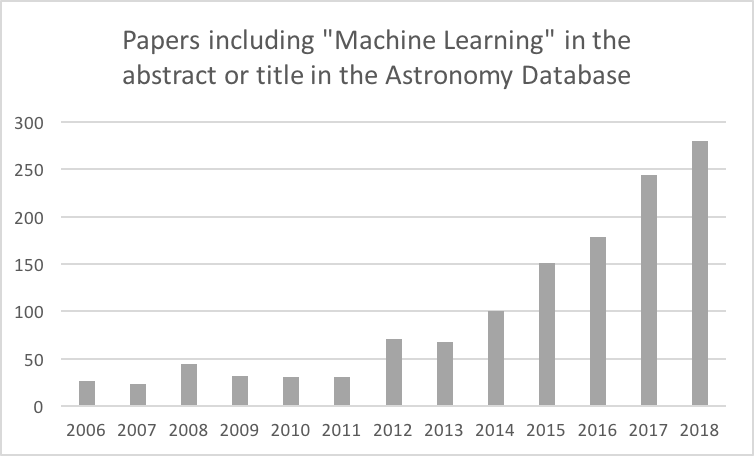}
\hspace{0.5cm}
\includegraphics[width=0.42 \textwidth]{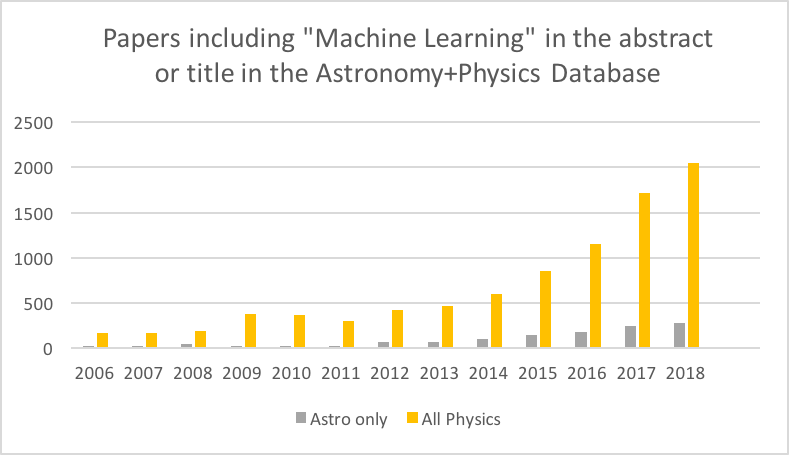}
 \caption{\small Machine learning related papers on the NASA/ADS archive from 2006 to today.}
   \label{Fig:papers}
   \vspace*{-0.2 cm}
\end{center}
\end{figure}

Machine learning and data mining methods have become ubiquitous in all fields of science, including Astronomy, in the last decade. Figure \ref{Fig:papers} shows the result of a search on the NASA-ADS archive service for papers that contain the words ``Machine Learning" either in the title or in the abstract, in the last 12 years, in the Astronomy and Physics archives; the growth trend is obvious. But of course it's not just papers; during this time, we have seen the emergence of dedicated conferences and sessions in conferences (such as this one), scientific associations, perhaps best described in the Astrostatistics and Astroinformatics Portal (ASAIP) \href{https://asaip.psu.edu/}{website}, special initiatives (for example the \href{https://cosmostatistics-initiative.org/}{COsmo INitiative}), dedicated programming packages, such as AstroPy \citep{Astropy2013,Astropy2018} and AstroML \citep{AstroML}, books, ad-hoc review panels from the National Science Foundation, and, of course, jobs. The most recent ASAIP blog post (10/2018) estimates that 15-20\% of job ads on the AAS job register have a strong component in statistics, machine learning, or data science. But what is machine learning? And most importantly, what can we do with it? \\[-0.5cm]

\section{What can we do with ML?} Machine learning is the art of teaching a machine to make informed decisions. Examples of its applications include recognizing and characterizing objects based on similarities and differences, picking up patterns, and distinguishing signal from noise. The following is a (certainly incomplete) list of important applications of machine learning methods to Astronomy. \\

\begin{figure}
\begin{center}
\includegraphics[width=0.8\textwidth]{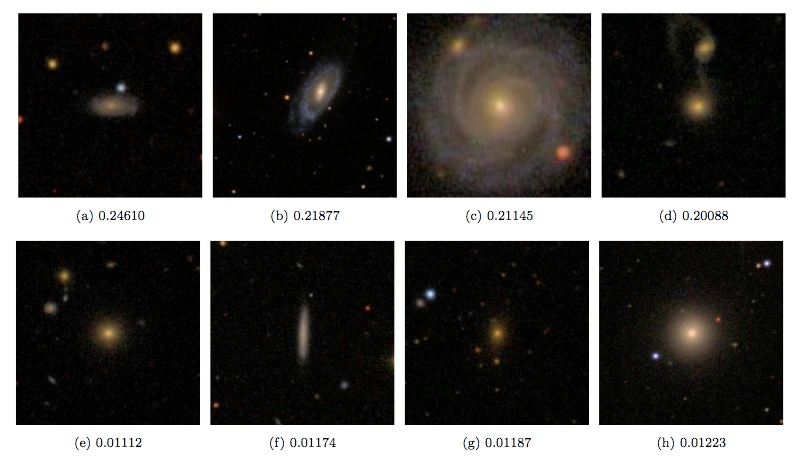}
 \caption{\small \cite{Dieleman2015} tackled the problem of galaxy morphology classification using deep learning.}
\label{Fig:Dieleman}
\end{center}
\end{figure}

 \noindent {\bf Save time.} This is perhaps the chief application of machine learning in a Universe as vast as ours. A typical example is the classification of galaxies according to their morphological type; \cite{delaCalleja2004} is, to my knowledge, the first example, but there have been many more since then (see \eg \href{http://adsabs.harvard.edu/cgi-bin/nph-ref_query?bibcode=2004MNRAS.349...87D&amp;refs=CITATIONS&amp;db_key=AST}{this link}); figure \ref{Fig:Dieleman} above shows a recent example from \cite{Dieleman2015}. In this case, trained humans are the best classifiers; we have an uncanny ability to generalize just from a few examples, and to weed out unimportant features. However, when the data sets become large (millions of objects), there are only two solutions: deploy the power of citizen science, with successful programs such as Galaxy Zoo \citep{Lintott2008}, or use machine learning. In a supervised approach, one would ask expert classifiers to provide ``solutions" (estimated galaxy types) for a subset of objects, the learning set, and then teach the machine to learn from those examples; in an unsupervised approach, the machine determines its own set of galaxy types directly from the data, then assigns each object to a class \citep{Hocking2018}. \\
    
   \noindent{\bf Provide a data-driven alternative to simplistic models.} The alternative to machine learning is usually a model-based approach in which one creates intuition-based models of the objects, typically dependent on some input variables that are the parameters of the model. For example, when modeling galaxies, we use stellar population models paired with parameters such as stellar mass, stellar assembly history, metallicity, dust \begin{wrapfigure}{l}{0.5\textwidth}
\vspace*{-0.3 cm}
\begin{center}
\includegraphics[width=0.45\textwidth]{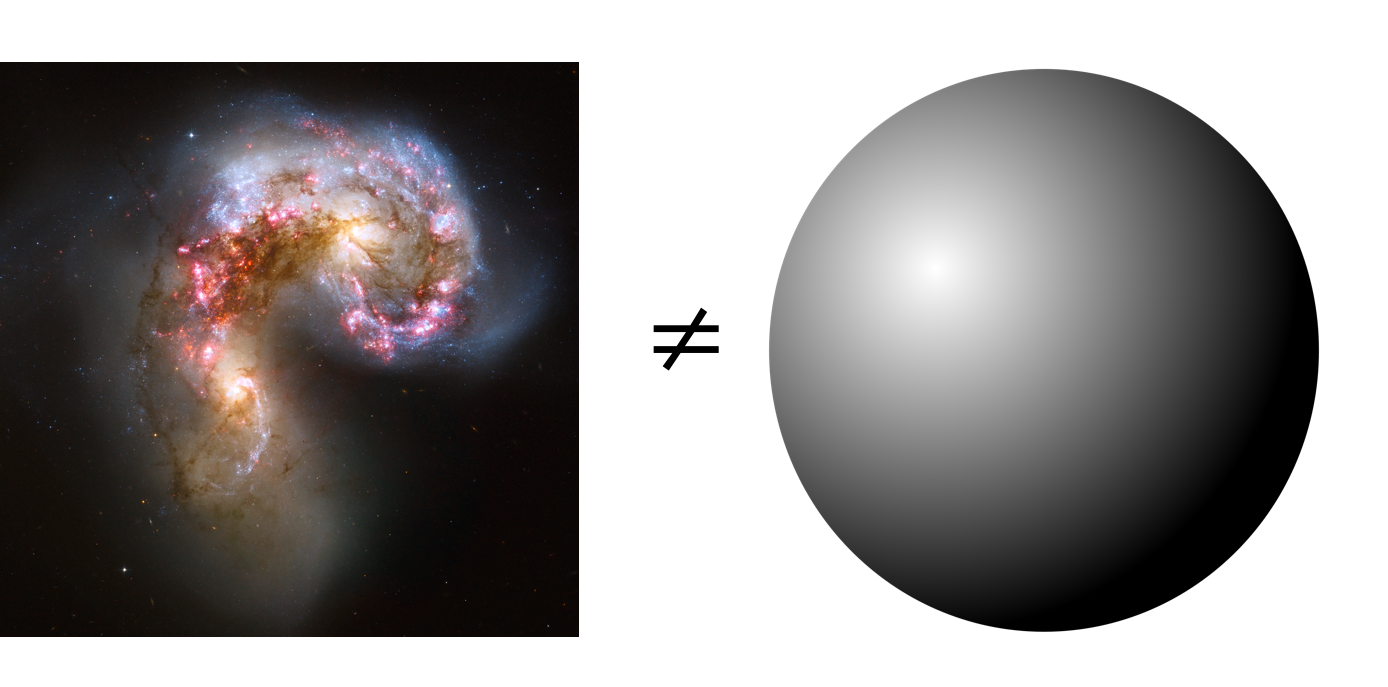}
 \caption{\small Alas, galaxies are not spheres.}
   \label{Fig:GalsNotSpheres}
   \vspace{-0.7cm}
\end{center}
\end{wrapfigure} abundance and properties, and so on. The models are obviously an extreme simplification of complex objects such as real galaxies (in my own cheesy words: galaxies are not spheres, see Fig. \ref{Fig:GalsNotSpheres}). As a result, when we build likelihood functions and we infer probability distribution functions for our parameters, we bias our results by imposing our choice of parameters and priors. This model-related bias is something that machine learning can hope to circumvent, since the relation between input variables (observables) and output (measurements, or predictions) is purely data-driven. \\

    \noindent{\bf Make problems more tractable.} This is another field in which machine learning algorithms are quite amazing. One very simple but incredibly helpful application is dimensionality reduction, the process of identifying patterns in the data that allow one to reduce the size of the data space, making data easier to visualize, manipulate, and analyze, as shown 
    \begin{wrapfigure}{r}{0.5\textwidth}
    \vspace*{-0.4 cm}
    \begin{center}
\includegraphics[width=0.45\textwidth]{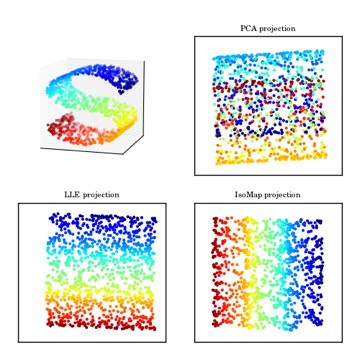}
 \caption{\small An example of 3D $\rightarrow$ 2D dimensionality reduction, with various degrees of information loss. Figure from \href{https://scikit-learn.org/stable/}{the sklearn website}.}

\label{Fig:trac}
\end{center}
\end{wrapfigure}
    in Fig. \ref{Fig:trac}. 
    Another great aspect is the increased resilience of ML methods to heterogeneous and sparse data. To stay within the regime of SED fitting, it is hard to create joint probability distributions to combine photometric and spectroscopic data, while for many ML methods, this problem just amounts to creating additional features. \\
    
      \noindent{\bf Facilitate serendipitous discoveries.}
 In 2017, a workshop called ``Detecting the Unexpected" took place at the Space Telescope Science Institute. One of the most interesting talks (at least for me, following from afar) introduced me to the work of \cite{Norris2017}, who looked at the ten most important discoveries of the Hubble Space Telescope and whether or not they were ``planned". The result of this analysis is shown below in Fig. \ref{Fig:DTU}. This - like many other examples - shows that one of the most important ways to prepare ourselves for further discoveries is to plan for what we don't know yet. In this respect, machine learning and data mining methods are really powerful; ML algorithms for pattern recognition and outlier detection are in general superior to model fitting, because the latter assumes, by construction, that we know what we are looking for.      
\begin{figure}[b]    
\begin{center}
\includegraphics[width=0.9\textwidth]{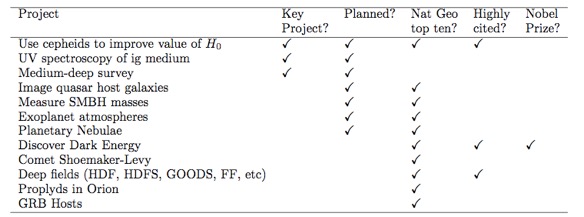}
 \caption{\small The analysis of Norris 2017 on the unexpected discoveries of the HST.}
\label{Fig:DTU}
\end{center}
\end{figure}

\section{Machine Learning versus Model Fitting: Who wore it better?}

While I do, in many ways, ``believe the hype" about machine learning, and I think it is important for young and old scientists alike to become proficient with these methods, I think it is important to realize that these techniques are just another set of tools in an astronomer/data scientist's skill set. I've often joked that machine learning methods are the ``Numerical Recipes" of this decade. They enable us to do some science that would be difficult to do otherwise; but as always, it's part of a scientist's job to understand how to ``attack" a difficult problem, with or without machine learning. As a general and incomplete rule, I think that when we understand the physics but we don't have data, we use model fitting, and when we don't understand the physics but we have data, we use machine learning; the figure \ref{Fig:MLvsMF} below summarizes some more of my thoughts on either approach, but the conclusion is that it's the synergy between these two approaches that is most powerful, because it gives us a strong cross-check of our problem-solving architecture. 

\begin{figure}[h]
\begin{center}
\includegraphics[width=0.8\textwidth]{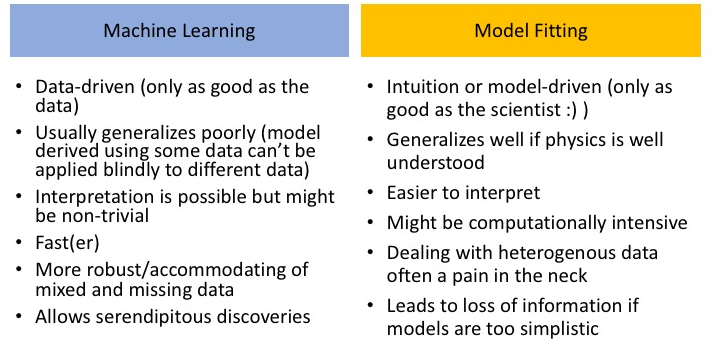}

 \caption{\small Some advantages and disadvantages of the two approaches. My claim is that {\bf synergy} is often the best strategy.}
   \label{Fig:MLvsMF}
  \vspace*{-0.7 cm} 
\end{center}
\end{figure}

\section{Why now?}

\begin{figure}[b]
\begin{center}
\includegraphics[width=\textwidth]{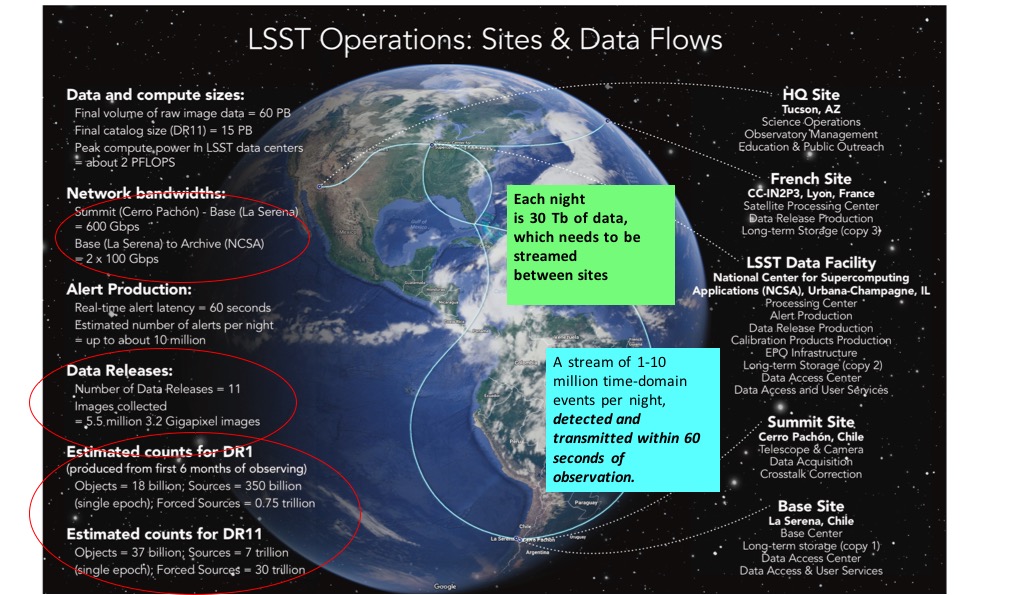}
 \caption{\small The ``Big Data" problem of LSST. From the LSST website; annotations are mine.}
   \label{Fig:LSST}
\end{center}
\end{figure}

I hope that the discussion above provides a fair assessment of why machine learning methods are useful in Astronomy. But another question arises about timing: after all, none of these problems or methods are new, so why have we seen such a strong increase in ML-related science in the last few years? The key to that, in my opinion, is that we have a {\it super} data problem (my way of saying it's not just {\it big} data). Sure, data are becoming bigger; if we use etendue as zeroth-order tracer of the data volume generated by a survey, we can see in the left panel of Fig. \ref{Fig:LSST} that the Large Synoptic Survey Telescope (LSST) dwarfs other surveys by a factor of 100 or more. However, big data alone would be largely tractable with more computing power. My claim is that we also have {\it better} data, because of the extended wavelength coverage, depth, and resolution of galaxy surveys. We have a {\it wide} data problem (to borrow a term I first heard from Alyssa Goodman at a 2015 talk), where we have multiple representations of the same object, contributing information from different parts of the spectrum. And we have a {\it new} data problem, where we are bound to discover things that we didn't think could exist, like hot Jupyters or super massive stars, or to unveil relationships that we don't yet understand. It is in dealing with wide and new data that, in my opinion, machine learning methods really shine, and the timing is now because of upcoming observatories like LSST, the James Webb Space Telescope, and Euclid.

\begin{figure}[h]
\begin{center}
\includegraphics[width=\textwidth]{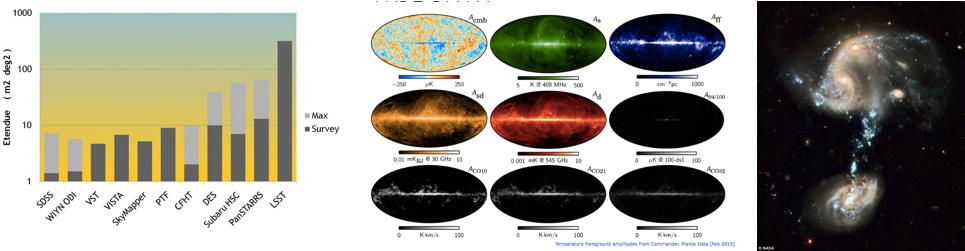}
 \caption{\small An exciting and terrifying scenario: we have bigger data, better data, wide data, and new data. Figure sources, from left to right: LSST website, Alyssa Goodman's 2015 NOAO talk, and the Hubble website.}
   \label{Fig:BigWideNew}
\end{center}
\end{figure}

\section{The example of LSST.} The Large Synoptic Survey Telescope provides a great example of the type of big data problem that astronomers are facing, and it's right around the corner, with a projected start date of 2023 (and an impressive record of being on schedule). Even the first of eleven data releases (DRs), anticipated to happen just six months after the beginning of science operations, will contain data for {\it eighteen billion} galaxies; DR11 will contain data for 37 billion galaxies, and will amount to the equivalent of 5.5 billion Gigapixel-size images. That's a lot of huge pictures! The problem of transferring and storing these data is only exacerbated by the fact that each night's observations will amount to 30 Tb of data; within those, 1-10 million time-domain events will need to be streamed within 60 seconds, because real-time detection of unusual and serendipitous phenomena is essential. In one word: Ouch!

\section{What's next?} This section summarizes my thoughts about what are going to be the most pressing issues that we will face in the next five years or so, and what algorithms and techniques will, in my opinion, become most important in the ``astronomy x data science" world. Of course, it's a limited, biased view - if I were a ML algorithm, I would have low precision and low recall! \\

\noindent
{\bf 5.1. Too much information, too little space.} One of the big issues we will face is going to be data storage. On the one hand, this applies to the raw data - for example, while searching and downloading the SDSS archive is a relatively manageable task for anyone with a laptop and a decent internet connection, the corpus of LSST data will certainly make this impossible. As a result, data compression, representation, and visualization techniques will become increasingly important. Examples of them range from the simple Principal Component Analysis to nonlinear techniques such as Self Organized Maps, t-distributed Stochastic Neighbor Embedding, Auto Encoders, and many others, some of which are described \href{https://scikit-learn.org/stable/modules/manifold.html}{here}.

\begin{figure}[t]
\begin{center}
\includegraphics[width=0.9\textwidth]{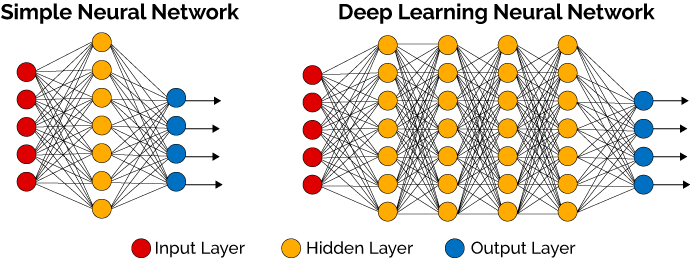}
 \caption{\small Deep Learning algorithms (right) use multiple layers to characterize the relation between input and output. Figure from \href{https://towardsdatascience.com/mnist-vs-mnist-how-i-was-able-to-speed-up-my-deep-learning-11c0787e6935}{this post}.}
   \label{Fig:DL}
\end{center}
\end{figure}

However, the tricky part is that different science applications will need different types of data, so the information loss associated to any technique may be perfectly acceptable for some projects, but not for others. The problem becomes even more pressing if one also considers {\it derived} products: catalogs, estimated parameters with their probability distributions, and so on. I found it telling that in my field of galaxy SED fitting, we spent a good fraction of the last decade arguing in favor of going beyond the best-fit model and using more sophisticated methods, such as various flavors of Markov Chain Monte Carlo methods, to explore the parameter spaces and generate full probability distributions. Instead, by the last CANDELS meeting (April 2018 at UC Riverside) the discussion had largely shifted towards what is the best way to make the products available, since full probability distributions are expensive to compute and store. I don't really have a proposed solution, but I argue that we will all collectively need to discuss what products should be made available, and to develop/learn smart tools for data representation. \\

\noindent
{\bf 5.2. Deep Learning FTW.} My prediction is that we will see a proliferation of deep learning methods (Fig. \ref{Fig:DL}), a subset of machine learning methods in which the input/output relation is modeled through a series of intermediate steps (layers) that make them really flexible and powerful in capturing features from raw data, moderating the need for feature selection and engineering, and modeling complex relationships.

The following are three types of neural networks that have already been shown to be very useful in Astronomy, and I expect that we'll hear a lot more about them in the next five years.

\begin{figure}[t]
\begin{center}
\hspace{0.8cm}
\includegraphics[width=0.9\textwidth]{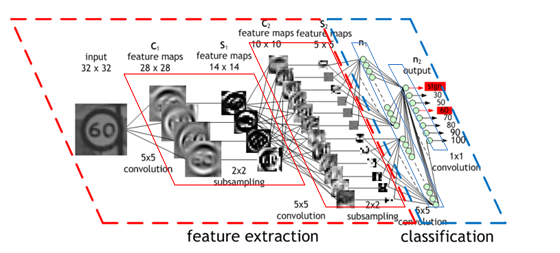} 
 \caption{\small A scheme of a convolutional neural network. The network includes convolution layers, in which complicated patterns are broken down into simple ones, creating data-driven features, pooling layers that are used for subsampling/dimensionality reduction, and a final fully connected layer that predicts the output. Figure from \href{http://parse.ele.tue.nl/mpeemen}{Maurice Peemen}.}
   \label{Fig:CNN}
\end{center}
\end{figure}

\underline{Convolutional Neural Networks.} Convolutional Neural Networks (CNNs) are growing in popularity in many areas of Astronomy, typically as a means of analyzing 2D image data (\textit{e.g.} \cite{Dieleman2015,tuccillo_2017, petrillo_finding_2017,Fussell2018,Huertas2018}). One powerful feature of CNNs is that they alternate different type of layers: convolution layers, used for feature extraction, and pooling layers, used for subsampling/dimensionality reduction. The final layer is a fully connected neural network, which generates the output. This structure, shown in Fig. \ref{Fig:CNN} above, enables them to autonomously extract significant features from images, reducing the need for feature selection and feature engineering, and making them suitable to process astronomical images, in which information is often present in the form of gradients, breaks, and spatial correlations. A very nice, detailed discussion of how CNNs can master morphology classifications can be found in this \href{http://benanne.github.io/2014/04/05/galaxy-zoo.html}{blog post}.

\underline{Recurrent Neural Networks.} Recurrent neural networks have the characteristic of using not just the inputs, but the output of previous queries as part of their output generation. This characteristic makes them very suitable for all sorts of time series (time domain) problems, in which we care about not just the characteristic of a single input instance but about a {\it sequence} of them. This is very helpful, among other cases, in outlier detection and hardware failure detection \citep{Naul2018,Zhang2018}.

\begin{figure}[h]
\begin{center}
\includegraphics[width=0.35 \textwidth]{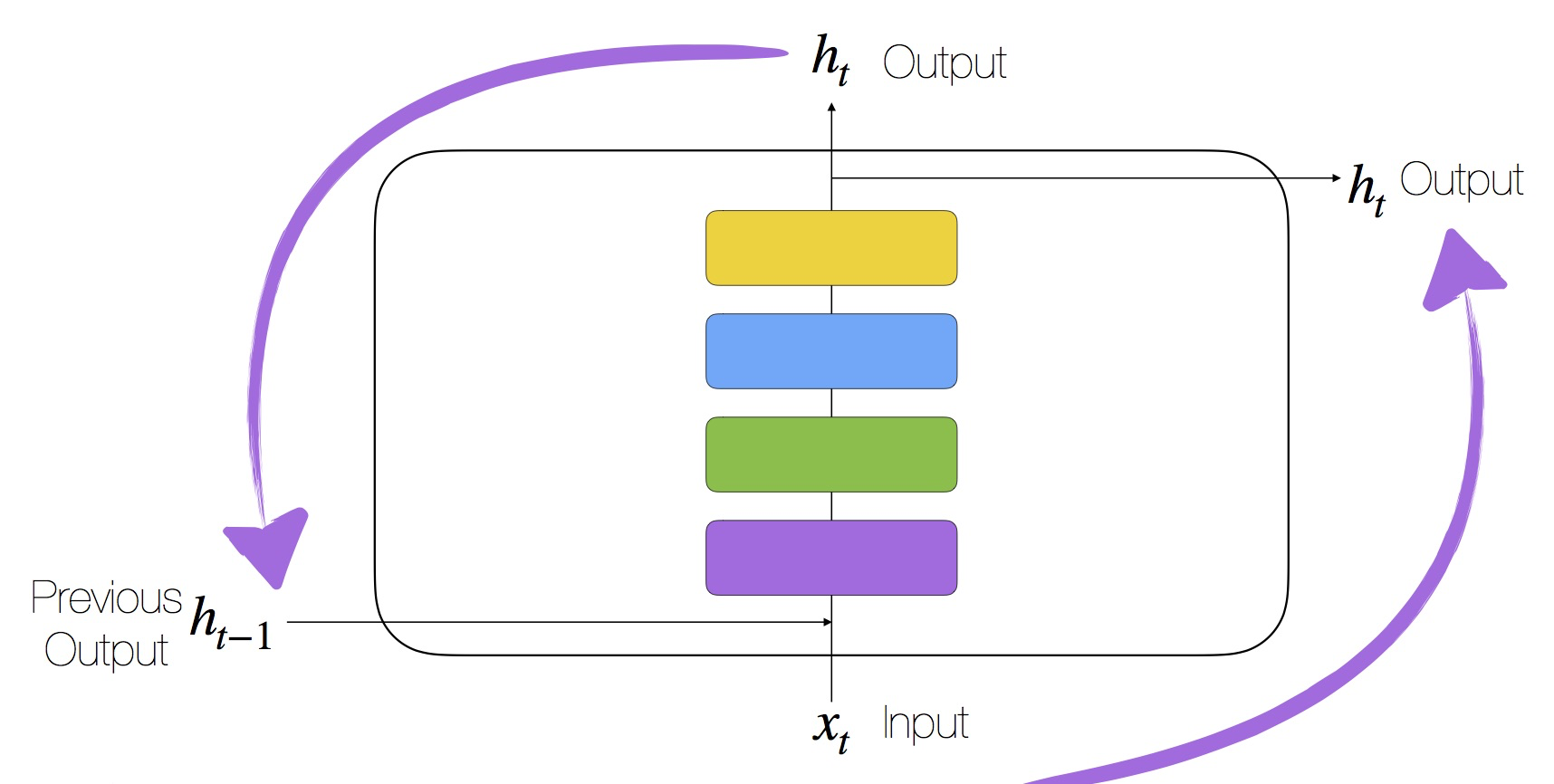}
\includegraphics[width=0.6 \textwidth]{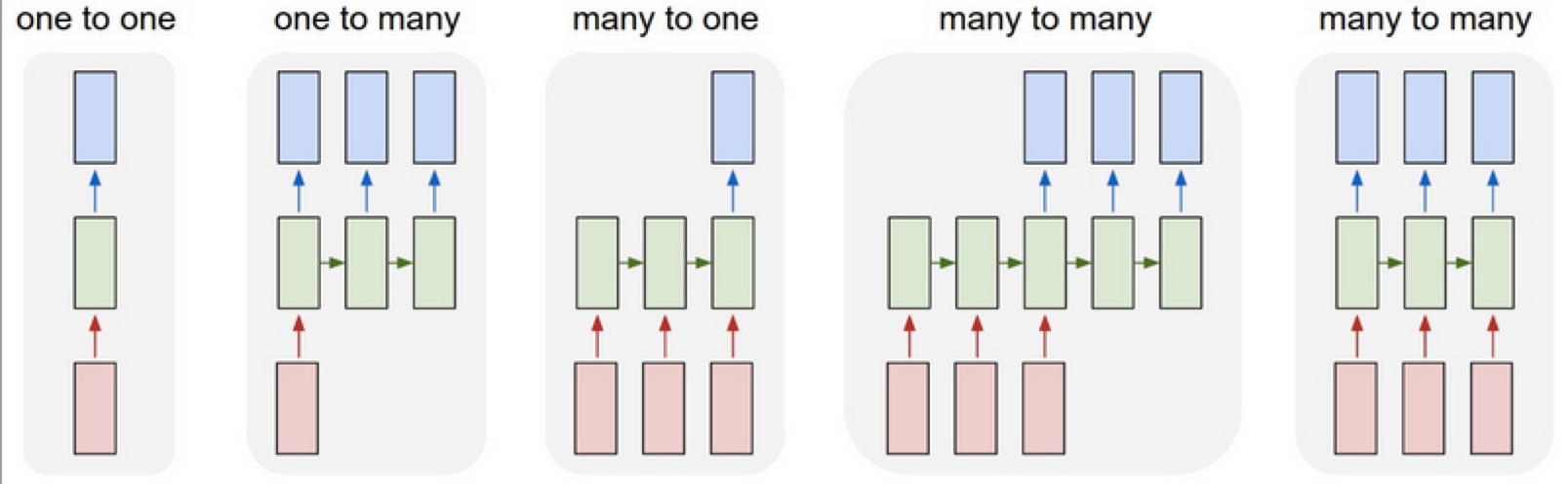}
 \caption{\small Left: In a Recurrent Neural Network, the information flow is not just forward-fed, but the unit processors of the network can use information about the output of the network in previous steps. Figure by \href{https://github.com/bmtgoncalves/}{Bruno Goncalves}). Right: Because of this features, RNNs are good at modeling relationships between series of inputs and series of outputs. Source: this \href{https://karpathy.github.io/2015/05/21/rnn-effectiveness}{blog post}.}
   \label{Fig:RNN}
\end{center}
\end{figure}
\newpage

\begin{figure}[t]
\begin{center}
\hspace{0.6cm}
\includegraphics[width=0.8\textwidth]{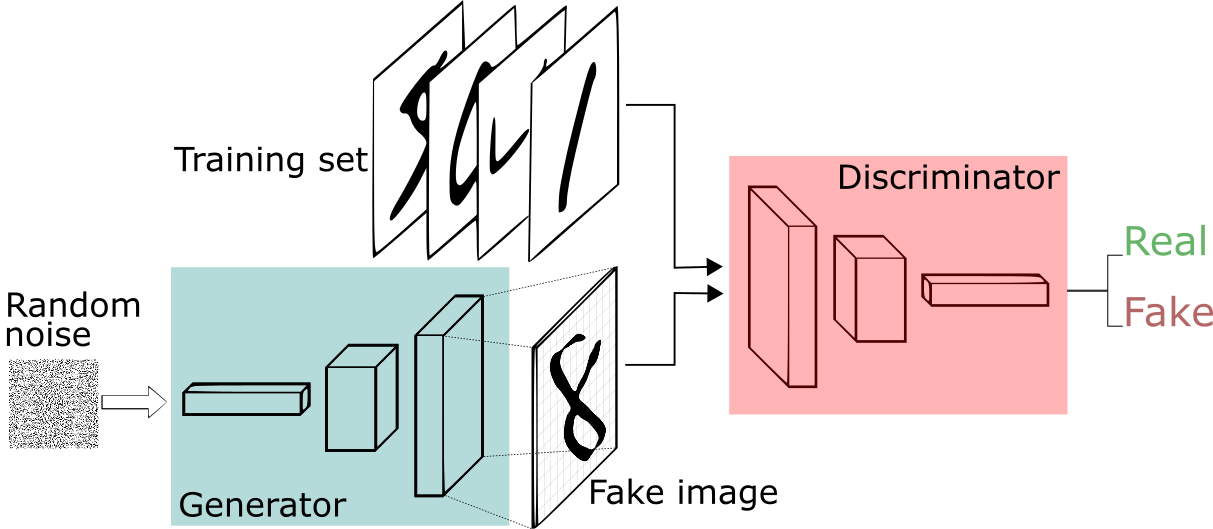} 
\newline
 \vspace{0.6cm} 
\includegraphics[width = \textwidth]{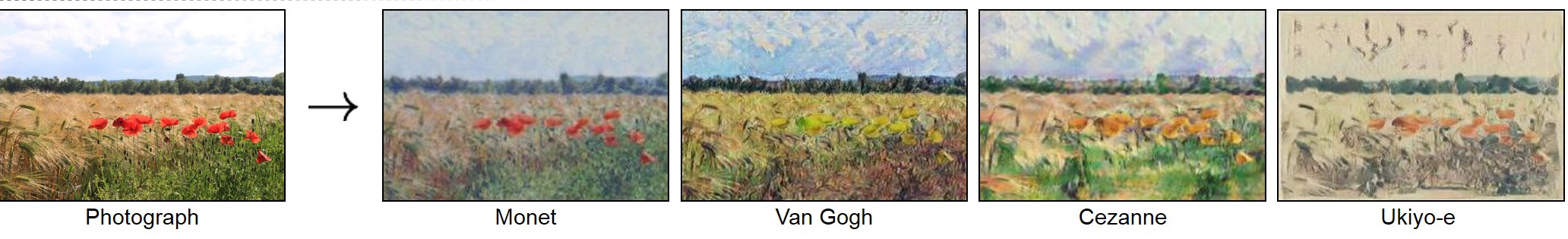}
 \caption{\small Top: Generative Adversarial Networks (GANs) learn how to generate realistic images thanks to their dual generator/discriminator structure. From \href{https://skymind.ai/wiki/generative-adversarial-network-gan}{Skymind}. Bottom: an example of style transfer, where the GAN processes the input photograph to generate art in the style of four different painters. From \cite{CycleGAN2017}.}
   \label{Fig:GAN}
\end{center}
\end{figure}

\begin{figure}[t]
\begin{center}
\includegraphics[width=0.52 \textwidth]{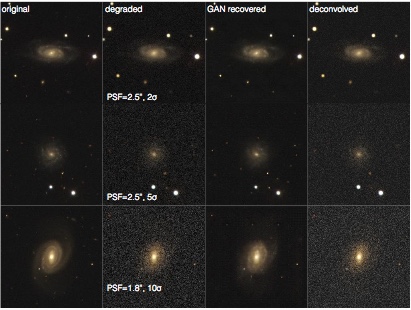}
\hspace{0.5cm}
\includegraphics[width = 0.41 \textwidth]{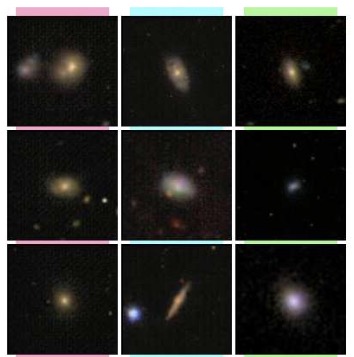}
 \caption{\small Left: An example of how GANs can be used to recover high-resolution galaxy images, from \cite{Shawinski2017} From left to right: the original SDSS image, the degraded image with a
worse PSF and higher noise level (indicated), the image as recovered by the GAN, and for comparison, the result of a deconvolution. Right: Generating simulated galaxy images with GANs; the first two columns show synthetic objects, the last one real SDSS images. From v1 of \cite{Fussell2018}.}
   \label{Fig:AstroGANs}
\end{center}
\end{figure}

\underline{Generative Adversarial Networks.} Generative Adversarial Networks \citep{Goodfellow2014} are built as a double system where one network attempts to build realistic examples of the objects, and the other one tries to detect whether they are fake or real. While the first applications were fraud detection, GANs have now been applied to a variety of problems, for example generating art in the style of a specific painter \citep{CycleGAN2017}; see Fig. \ref{Fig:GAN} above. This \href{https://medium.com/@jonathan_hui/gan-some-cool-applications-of-gans-4c9ecca35900
}{blog post} has a wonderful collection of examples. In Astronomy, GANs' recent claims to fame have been ``upsampling" image resolution \citep{Ledig2016,Shawinski2017}, and generating synthetic galaxy images that look realistic at the object and at the population level \citep{Fussell2018}, as shown in Fig. \ref{Fig:AstroGANs}. Both of these approaches have profound consequences for Astronomy, because they potentially lead to reduced observing times and to scaling down the processing time for large galaxy evolution simulations. \\

\noindent
{\bf 5.3. (Sorry,) We need to talk to non-astronomers.} An important consequence of the widespread impact of machine learning techniques in Astronomy is the fact that we need to develop interdisciplinary collaborations - for real! While this is in general always a good idea, I finally see it happening and it's in part due to the fact that data science seems to provide a common language for astronomers, statisticians, mathematicians, and computer scientists. One positive example is the fact that there have been several Astronomy/Physics challenges on the \href{https://www.kaggle.com/}{Kaggle website}, from the 2014 Galaxy Zoo challenge and Higgs boson challenge to recent ones like the \href{https://www.kaggle.com/c/PLAsTiCC-2018}{PLASTICC challenge} for light curves classification in LSST data and the ongoing \href{https://www.kaggle.com/c/LANL-Earthquake-Prediction}{earthquake prediction challenge} promoted by Los Alamos National Lab. Personally, I think this is a wonderful thing. \\ 

\noindent
{\bf 5.4. What the heck are we doing? Model interpretability/Explainability.} One drawback of powerful deep learning methods (but also of simpler models with heavily engineered features) is that they are difficult to interpret, hindering the process of producing generalizable insights. As a result, the last few years have seen the advent of interpretable machine learning (\eg \citealt{Molnar2018}), a discipline that seeks to increase the transparency of the machine's decision-making process. Useful methodology tools for this task are LIME (Locally Interpretable Model-Agnostic Explanations, \citealt{Ribeiro2016}), which uses surrogate simpler models to map the input-output relationships of an "opaque" algorithm such as a CNN; SHAP (SHapley Additive exPlanations, \citealt{Lundberg2018}), which uses game theory to understand the interactions between different features (players) and their roles in the final decision, and Recourse Analysis \citealt{Ustun2018}, which aims to build interpretability by understanding what feature modifications would change the outcome of a learning algorithm. The topic is controversial in the computer science community, and I believe that the trade-off between predictive power and interpretability should be decided on a discipline-by-discipline, or even case-by-case basis; see \cite{Lipton2016} for a great discussion. Nonetheless, I imagine that many of these will be helpful in our quest to further our understanding of the physics of galaxies. \\[-0.5cm]

\section{Conclusions}

Machine learning and data mining methods provide an exciting opportunity to deal with our ``super data" problems. Here are some ideas on how to leverage them: \\

\begin{itemize}

\item {\bf Teach/learn this stuff early and often}. There is no doubt that any Astronomy graduate student can benefit from learning about data science methods, and there are infinite possibilities to do so. Many resources, such as classes and books, are available for free online; there are also dedicated initiatives such as the \href{https://www.lsstcorporation.org/fellowship_program}{LSST data science fellowship}. \\[-0.2cm]

\item {\bf (Encourage mentees to) Code in Python.} This is essential, I think, for two reasons: one is convenience, because so many of the dedicated packages and support networks are available for Python, and second, because learning Python (and data science tools) will keep options open in case a career change is desired or necessary. I should add that some database experience and distributed computing experience is also quite desirable for this purpose. \\[-0.2cm]

\item {\bf (Encourage mentees to) Talk to people and go to conferences outside your field}. This year, the most inspiring talks I have heard were given in non-academic settings. I want to acknowledge these people here: Mikio Braun (Zalando) and Adil Aijaz (Split Software) at Strata Data NYC, Manojit Nandi (Rocketrip) at PyGotham, Stephanie Yang (Foursquare) at a NYC Women in Machine Learning and Data Science meetup, and Francesca Lazzeri (Microsoft) at DataX NYC. This is of course driven by my own interests, but I think it's generally true that hearing about different topics and gathering fresh perspectives is exhilarating, and I have found these environments generally welcoming (not to mention the networking value). \\[-0.2cm]

\item {\bf (Encourage mentees to) Publish your code and make your results reproducible}. As the use of computational techniques becomes increasingly essential and algorithms become possibly more opaque, sharing code, even if it's not pretty, is the main way of keeping each other accountable and learning from each other. Plus, having an online presence on github is a good idea in view of career advancement, both in academia and in industry. \\[-0.2cm]
 
\item {\bf Rest assured that science is science.} Machine learning is not a magic wand; it's just a way of enhancing the tool set that we have. In fact, I think that it is even more essential to learn how to formalize, attack, solve, and test our solutions to problems now that we have more ways to solve them. In other words, the scientific method stands high and strong, and intuition and rigor continue, even more than before, to be the pillars of advancement in  science. \\[-0.8cm]

\end{itemize}

\end{document}